\documentclass[runningheads]{llncs}
\usepackage{amsmath}
\usepackage{amssymb}
\usepackage{enumerate}
\usepackage{algorithm}  
\usepackage{algorithmic}  
\usepackage{MnSymbol}
\usepackage{multirow}
\usepackage{subcaption}
\usepackage{graphicx}
\usepackage{times}  
\usepackage{helvet}  
\usepackage{courier}  
\usepackage[hyphens]{url}  
\usepackage{graphicx} 
\usepackage{caption} 
\usepackage{algorithm}
\usepackage{algorithmic}
\usepackage{xspace}
\usepackage{amssymb}
\usepackage{booktabs}
\usepackage{amsmath}
\begin{document}
\title{MOT: A Mixture of Actors Reinforcement Learning Method by Optimal Transport for Algorithmic Trading}
\titlerunning{MOT: A Mixture of Actors Reinforcement Learning Method}
\newcommand{\themodel}{MOT\xspace}
%

%
\author{Xi Cheng\inst{1,2} \and
Jinghao Zhang\inst{1,2} \and
Yunan Zeng\inst{1,2} \and
Wenfang Xue\inst{1,2}}
\authorrunning{X. Cheng et al.}
%
\institute{
School of Artificial Intelligence, University of Chinese Academy of Sciences \and Institute of Automation, Chinese Academy of Sciences\\
\email{\{xi.cheng, jinghao.zhang, yunan.zeng\}@cripac.ia.ac.cn, wenfang.xue@ia.ac.cn}}

\maketitle              
\begin{abstract}
  Algorithmic trading refers to executing buy and sell orders for specific assets based on automatically identified trading opportunities. 
  Strategies based on reinforcement learning (RL) have demonstrated remarkable capabilities in addressing algorithmic trading problems. However, 
  the trading patterns differ among market conditions due to shifted distribution data. Ignoring multiple patterns in the data will undermine the performance of RL. 
  In this paper, we propose \themodel, which designs multiple actors with disentangled representation learning to model the different patterns of the market. Furthermore, we incorporate the Optimal Transport (OT) algorithm to allocate samples to the appropriate actor by introducing a regularization loss term. 
  Additionally, we propose Pretrain Module to facilitate imitation learning by aligning the outputs of actors with expert strategy and better balance the exploration and exploitation of RL.
  Experimental results on real futures market data demonstrate that \themodel exhibits excellent profit capabilities while balancing risks.  
  Ablation studies validate the effectiveness of the components of \themodel.

\end{abstract}
\section{Introduction}
The goal of algorithmic trading is to maximize long-term profits while keeping risks within an acceptable range \cite{pricope2021deep}. Compared to the traditional approach of relying on the expert judgment of trading timing, algorithmic trading is highly automated and efficient.

Traditional technical analysis methods include mean reversion \cite{jegadeesh1993returns}, momentum investing \cite{jegadeesh2002cross}, multi-factor models \cite{fama1996multifactor}, etc. However, financial market data is non-stationary with a low signal-to-noise ratio.
Expert-designed technical analysis methods can't generate profits under diverse market conditions.
Deep learning methods excel at capturing intricate price patterns and enhance models' performance~\cite{xu2021hist,xu2021rest,lin2021learning}. 
However, the process from supervised models' output to actual investment still requires the construction of strategies, which introduces expert knowledge and subjectivity. RL methods don't require carefully designed strategies by humans. They take market information as states and output trading decisions directly, which makes it easy to incorporate the unique financial constraints (e.g. transaction costs and slippage) into environments. RL has achieved SOTA in many quantitative investment tasks~\cite{liu2020adaptive,de2020tabular,yuan2020using}.

\begin{figure}[t]
  \centering
  \includegraphics[width=\linewidth]{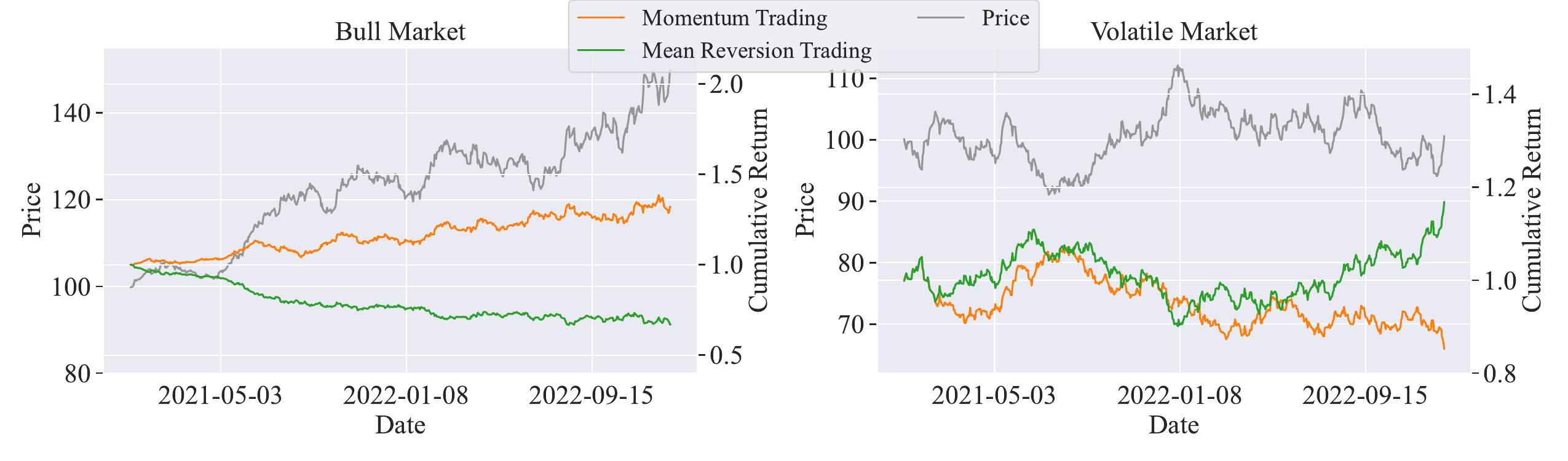}
  \caption{Profit of strategies in different market conditions. A bull market is suitable for momentum trading, while a volatile market is suitable for mean reversion trading.}
  \label{fig:marketStrategy}
\end{figure}
However, these methods rely on the assumption that financial data always follow the same distribution. Data patterns often switch in real scenarios. E.g. the most common way to classify market patterns is into two categories: stable (momentum) and volatile (reversal) markets, which require two categories of strategies \cite{li2018machine}. These two phenomena are not independent but intertwined with each other.
As shown in Figure \ref{fig:marketStrategy}, when bullish forces $>$ bearish forces or are evenly balanced, the market is in a stable upward (bull) or a volatile state respectively. Momentum trading strategy models the momentum effect of stable market and mean reversion strategy models the reversal effect of volatile market.
The same strategy can yield significantly different returns in different market conditions. Inspired by a mixture of experts~\cite{fedus2022switch}, we propose \themodel, which models multiple actors with disentangled representation learning and extracts various pattern information in RL. To allocate samples to agents appropriately, we introduce the Allocation Module with Optimal Transport (OT) regularization loss.

Previous research \cite{liu2020adaptive} has introduced imitation learning to RL, allowing agents to learn information from expert knowledge. However, in the early stages of imitation learning, the sampled action used in the training process is not from the agent's generation but is directly given by the expert which is stored in the buffer. As a result, the true output action of the agent differs significantly from the action stored in the buffer. To solve this problem, \themodel introduces a pre-training method based on supervised learning to imitation learning. We expect the output generated by the agent to be closer to the expert strategy in imitation learning so the model can be initialized in a better stage.

The training process of \themodel can be divided into three stages: first, the Pretrain Module uses supervised learning to train only the actor with expert strategy. Then we use the expert strategy to fill the buffer and train the RL model by imitation learning. After that, \themodel uses multiple actors to model different market patterns and uses OT to solve the problem of pattern allocation. The contributions are summarized as follows:

1) \themodel is the first that introduces OT algorithm to RL for mining various trading patterns. Allocation Module allocates different samples to appropriate actors. 

2) \themodel is also the first study that addresses the imitation learning gap between the actor's output and the buffer. \themodel introduces a supervised Pretrain Module before imitation learning, which allows the real actor's output to be closer to the expert strategy.

3) Experiments show \themodel has great profitability in different market modes while balancing risks. Further studies confirm the effectiveness of three components of \themodel.

\section{Problem Formulation}

\begin{table}[tb]
  \caption{Changes in Position Based on Trading Signals}
  \label{tab: Changes}
    \begin{tabular}{cccl|cccl}
    \toprule
    \textbf{Po}&\textbf{Action}&\textbf{Po\'}&\textbf{Operation}&\textbf{Po}&\textbf{Action}&\textbf{Po\'}&\textbf{Operation}\\
    \midrule
     0 & 1 & 1 & Take a long position&0 & -1 & -1 & Take a short position\\
     1 & 1 & 1 & No operation&-1& -1 & -1 & No operation\\
     -1& 1 & 1 & Close the position then go long&1 & -1 & -1 & Close the position then go short\\
  \bottomrule
\end{tabular}
\end{table}

The algorithmic trading problem can be represented as Markov Decision Process (MDP) $\mathcal{M} = \langle \mathcal{S, A, P, R,} \gamma \rangle$, where $\mathcal{S}$ represents the state space provided by the environment, $\mathcal{A}$ represents the action space, $\mathcal{P: S \times A \times S} \rightarrow [0, 1]$ is the probability function of the conditional state transitions, $\mathcal{R: S \times A} \rightarrow \mathbb{R}$ is the reward function, and $\mathcal{\gamma} \in (0, 1)$ is the discount factor. The specific definition of the five-tuple for MDP is as follows:

\textbf{The State space $\mathcal{S}$:} The state $\textbf{S}_t = [\textbf{S}_t^m;\textbf{S}_t^a ] \in \mathcal{S}$. The account indicators $\textbf{S}_t^a=[ s_t^{a_1}, s_t^{a_2}, ...]$ describe the trader's positions, account cash balance, margin, returns, and other related information of the trader's account.  $\textbf{P}_t=[ p_t^o, p_t^h, p_t^l, p_t^c, v_t^o, v_t^a ]$ represents the Opening-High-Low-Closing (OHLC) prices, trading volume, and trading value. $\textbf{Q}_t=[ q_t^1, q_t^2, ... , q_t^i]$ are derived from $\textbf{P}_t$ and technical analysis. The market indicators $\textbf{S}_t^m= [\textbf{P}_t;\textbf{Q}_t ]$ include the volume-price data $\textbf{P}_t$ and the technical indicators $\textbf{Q}_t$. 

\textbf{The Action space $\mathcal{A}$}: The action $a_t \in \left\{-1, 1\right\}$ represents the trading signal output by the agent. -1 corresponds to short selling and 1 corresponds to a long position. We define the agent to trade in units of contracts. The actual execution of trades depends on the trading signal and the trader's existing positions. The specific changes in position and action are summarized in Table \ref{tab: Changes}, where $Po$ means position. 

\textbf{The Transition Function $\mathcal{P}$}: We assume that the actions of individual traders do not affect the overall asset price in the market. This implies that the observation transition function of market indicators is independent of trading behavior, i.e. $\mathcal{P}(\textbf{S}_{t+1}^m | \textbf{S}_t) = \mathcal{P}(\textbf{S}_{t+1}^m | \textbf{S}_t, a_t)$. However, the observation transition function of account prices is influenced by trading behavior, i.e. $\mathcal{P}(\textbf{S}_{t+1}^a | \textbf{S}_t) \neq \mathcal{P}(\textbf{S}_{t+1}^a | \textbf{S}_t, a_t)$.

\textbf{The Reward $\mathcal{R}$}: We choose the closing price $p_t^c$ to calculate profit $r_t$. To better simulate real market, we set transaction fee rate $\mu$ \footnote{Transaction costs are charged as a percentage of the contract.} and slippage $\sigma$ \footnote{Slippage refers to the difference between the expected and the actual execution price.}. The profit $r_t$ is defined as $r_t=(p_t^c-p_{t-1}^c-2\sigma)\cdot a_{t-1}-\mu  \cdot p_t^c \cdot |\Delta po|$,
where $\Delta po =  Po' -Po$. 
When setting rewards, it is inappropriate to consider only the profit without taking into account the risk. 
The Sharpe ratio is the most widely used indicator for balancing risk and returns \cite{sharpe1966mutual}, defined as $SR=\frac{mean(r_t)}{std(r_t)}$.
To measure the impact of the profit on $SR$ each step, we adopt the Differential Sharpe Ratio (DSR) \cite{moody1998reinforcement} as the reward. Considering that the adjacent data is more important than distant previous data in algorithmic trading, DSR employs the smoothing technique of Exponential Moving Average (EMA). $DSR_t$ is defined as:
\begin{equation}
    \label{equ:r7}
  DSR_{t} = \frac{B_{t-1} \Delta A_{t} - \frac{1}{2} A_{t-1} \Delta B_t}{(B_{t-1} - A_{t-1}^2)^{\frac{3}{2}}}  ,
\end{equation}
representing the impact of each new profit $r_t$ on $SR$ after applying EMA. $A_t$ is the first moment and $B_t$ is the second moment of profits $r_t$ estimated by EMA. We utilize the $DSR_{t}$ as the reward $\mathcal{R}$. 
If the account money is insufficient, the trading will be terminated in advance, and we simulate this by setting a margin threshold. 

\begin{figure*}[tb]
  \includegraphics[width=\textwidth]{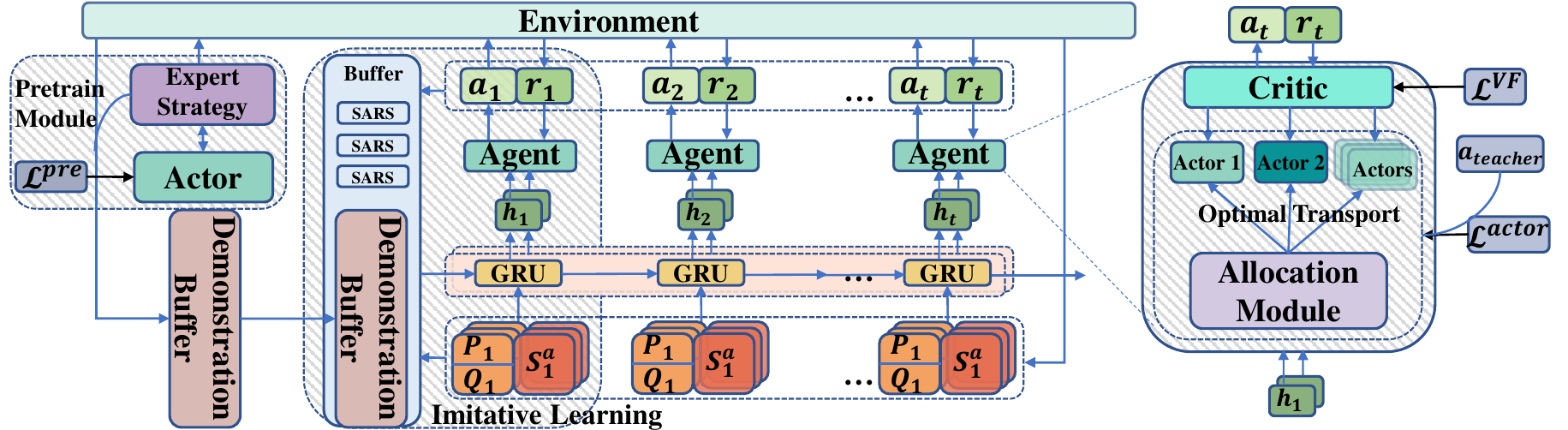}
  \caption{The architecture of \themodel. First, we pretrain the actor using the expert strategy and then proceed with imitation learning. We model different market patterns using multiple actors and allocate samples to the actors using the Allocation Module.}
  \label{fig:Architecture}
\end{figure*}
\section{Methodology}
The overview of \themodel is present in Figure \ref{fig:Architecture}. First, to ensure alignment between the actions in Demonstration Buffer and the actual outputs of the actor, we introduce Pretrain Module. Second, we leverage imitation learning to initialize the RL algorithm. Third, we use multiple actors with disentangled representation learning and model various market conditions. Last, Allocation Module allocates samples to different actors by OT algorithm. 

\subsection{Imitation Learning}
\label{41}
In RL-based algorithmic trading, the initial exploration phase is often inefficient and yields low profits. Imitation learning leverages expert knowledge and provides the actor with a favorable starting point.
We employ PPO \cite{schulman2017proximal} as the backbone to address the MDP problem. 
To capture the temporal patterns of states $\textbf{S}_t$, we utilize Gated Recurrent Units (GRU) \cite{chung2014empirical} to obtain the hidden representation $h_t=GRU(h_{t-1},\textbf{S}_t)$ of states $\textbf{S}_t$. 
$h_t$ is then fed into the actor and critic networks as inputs. 

The actor network aims to find the optimal policy $\pi$ by maximizing the advantage function. The input is the environment state $\textbf{S}_t$ and the output is the action $a_t$. To ensure sufficient exploration by the agent, we add noise $\varepsilon$ to the output of the actor network.
The actual executed action $a_t = \pi^\theta (h_t) + \varepsilon$, where $\varepsilon$ represents the noise, $\pi$ is the policy given by actor network with parameters $\theta$. The trading experience trajectories (SARS: state, action, reward, new state) are stored in the buffer $\mathcal{B}$. After sampling, we update the gradients of the actor network and the critic network using the data from $\mathcal{B}$.

The value function $V$, computed by the critic network with parameters $\omega$, estimates the value of the sample under state $\textbf{S}_t$. It is optimized through the loss function:
\begin{equation}
\label{eq: 9}
\mathcal{L}^{VF}(\omega) = \mathbb{E} \left[ (V_{\omega}(\textbf{S}_t) - V_t)^2 \right], 
\end{equation}
where $V_t=\sum_{t'=t}^{T-1} \mathbb{E}[ \gamma^{T-t'-1} DSR_{t'}(\textbf{S}_{t'}, a_{t'})]$ represents the empirical value of the accumulated future rewards $DSR$ and $T$ is the total number of time steps.

Let $\delta _t^V=DSR_t+\gamma V(\textbf{S}_{t+1}) - V(\textbf{S}_t)$ represent the advantage value estimation. In our research, the advantage function is computed by generalized advantage estimator (GAE) \cite{schulman2017proximal}: $\hat{A}_t^{GAE(\gamma, \lambda)} = \sum_{k=t}^{T-1} (\gamma \lambda)^{k-t} \delta_k^V,$
where $\gamma$ is the discount factor, $\lambda$ represents the trade-off between variance and bias.

PPO introduces a surrogate objective function to measure the similarity between the updated policy and the previous policy. The policy ratio formula is $\frac{\pi_{\theta}(a_t|\textbf{S}_t)}{\pi_{\theta_{\text{old}}}(a_t|\textbf{S}_t)}$. $\pi_{\theta_{\text{old}}}$ and $\pi_{\theta}$ represents the original and updated policy respectively.
The objective function $\mathcal{L}^{CLIP}(\theta)$ for policy update is as Equation \ref{eq: 11}, $\epsilon$ is the clipping threshold.

We employ the commonly used Dual Thrust \cite{kim2007hybrid} as the expert strategy to provide demonstration actions. We store the demonstration trajectory SARS in Demonstration Buffer (DB)
and train the agent using samples from DB.
The training of the actor-critic network in imitation learning follows the same approach as the PPO algorithm, with the only difference being that the training data is from DB. Subsequently, the actor-critic network continues to train by PPO method, as shown in Equation \ref{eq: 9} and Equation \ref{eq: 11}:
\begin{equation}
\label{eq: 11}
\mathcal{L}^{CLIP}(\theta) = \mathbb{E} \left[ \min ( \frac{\pi_{\theta}(a_t|\textbf{S}_t)}{\pi_{\theta_{old}}(a_t|\textbf{S}_t)} \hat{A}_t, clip(r_t(\theta), 1 - \epsilon, 1 + \epsilon) \hat{A}_t )\right].
\end{equation}

\subsection{Pretrain Module}\label{42}
The Pretrain Module is used to align the actions in the buffer $\mathcal{B}$ with the outputs of the actor. As mentioned before, it can be observed that $a_{expert}$ in DB is directly provided by Dual Thrust strategy rather than generated by the actor network $\pi_{\theta}$. Therefore, when using the demonstration data for gradient descent of the network, there is a significant discrepancy between the distribution of the actor network's output action $\pi_{\theta}$ and the action $a_{expert}$ \cite{liu2020adaptive}. This has a negative impact on the stability of the RL network. 

To address this issue, we aim to align the output action $a_t=\pi_{\theta}$ of the actor network with the expert-provided action $a_{expert}$ by training the actor network using supervised learning. 
The loss function is defined as Equation \ref{eq:12}:
\begin{equation}
\mathcal{L}^{pre} = CrossEntropy(a_{expert}, \pi_\theta (h_t))
\label{eq:12}
\end{equation}

Pretrain Module accelerates the actor's understanding of the task
by mimicking expert strategies 
and enhances the actor's ability to effectively engage in the imitation learning process. Pretrain Module is positioned before imitation learning as Figure \ref{fig:Architecture}.

\subsection{Multiple Actors}\label{43}
We employ multiple actors to model strategies in different patterns. Futures data is derived from the trading activities of numerous participants and reflects different trading patterns~\cite{ritter2003behavioral}. Ignoring multiple patterns will reduce the performance of models~\cite{houlsby2019parameter}. 
All $k$ actors of \themodel are constructed in the same manner, as depicted in Figure \ref{fig:Architecture} and Equation \ref{eq: 11}.
For convenience, we illustrate how the model is trained with $k=2$.

To integrate the outputs of the two actors, we use an Allocation Module to assign weights to them. Regarding the construction of the Allocation Module, we first consider what inputs should be provided to it. The historical sequence of futures states $\textbf{S}_t$ plays a significant role in determining the current market patterns. Additionally, the historical decision errors of different actors represent their decision-making performance and also influence the current sample allocation. We use GRU to extract latent feature representations from $\textbf{S}_t^i$, denoted as $\hat{h}_t^i=GRU(\hat{h}_{t-1}^i,\textbf{S}_t^i)$, where $i$ means i-th sample. As the calculation of sample decision errors, we provide posterior teacher actions on the training set. The teacher action $a_{teacher}=1$ when the futures close price $p_t^c$ increases in the next time step and $-1$ otherwise. Let $a^{i1}$ and $a^{i2}$ represent the action output by actor 1 and actor 2. The sample decision error $\textbf{e}_t^i$ is then computed as $[a_{teacher\ t}^i-a_t^{i1}, a_{teacher\ t}^i-a_t^{i2}]$. To avoid introducing future information, we utilize the previous error $\textbf{e}_{t-1}^i$. Subsequently, we concatenate $\hat{h}_t^i$ and embedding of error sequence $\textbf{d}_{t-1}^i= GRU(\textbf{d}_{t-2}^i,\textbf{e}_{t-1}^i)$ and feed them into a fully connected layer to predict the allocation results, denoted as $\textbf{b}_t^i = FC(\hat{h}_t^i, \textbf{d}_{t-1}^i)$. 
\begin{figure}[bt]
  \centering
  \includegraphics[width=\linewidth]{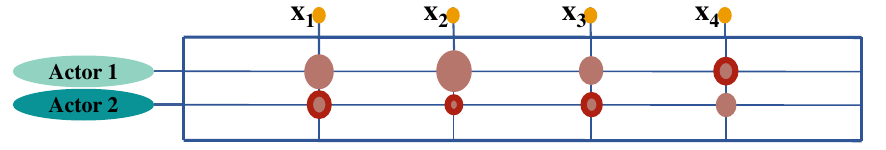}
  \caption{OT refers to assigning $x$ to the actor with the minimum $L_{err}^{ij}$ while achieving a balanced allocation proportion, $\frac{x\ to\ Actor\ 1}{x\ to\ Actor\ 2}\approx\frac{w_1}{w_2}$.
  The pink circles represent $L_{err}^{ij}$.}
  \label{fig: OT}
\end{figure}
In different patterns Allocation Module should have different attention for the two actors in Equation \ref{eq: 13}, 
where $q_t^i$ represents the allocation weights, and $a_t^i$ represents the final action. To ensure the discrete differentiability of the Allocation Module, we utilize the gumbel-softmax method \cite{jang2016categorical} to compute Equation \ref{eq: 13}. It is worth noting that the allocation of samples is not binary, but rather a soft allocation ranging $0<\textbf{q}_t^{i}<1$.
\begin{equation}
\label{eq: 13}
    \textbf{q}_t^i = softmax(\textbf{b}_t^i) , \ a_t^i = {\textbf{q}_t^{i}}^T [a_t^{i1}, a_t^{i2}],
\end{equation}

However, if the actors want to learn different patterns, the representations should be as dissimilar as possible. Inspired by disentangled representation learning, we take the inputs $x$ of the actors' last layers as the representations and design a disentangled loss to enable the agent to learn different patterns, $\mathcal{L}^{dis}=\sum_{i=1}^{N} x_{i1} \cdot x_{i2}$.

\begin{algorithm}[tbp]
\caption{Training process of {\themodel}}
\begin{algorithmic}[1]%
    \STATE Initialize actor network parameters $\theta_0$, critic network parameters $\omega_0$ and epochs $K$
    \STATE Obtain the expert strategy
    \STATE Pretrain the actor by $\mathcal{L}^{pre}$ in Equation \ref{eq:12}
    \STATE Add the expert strategy to DB and train by imitation learning, get the dual policies $\pi_{\theta_j}(a|\textbf{S})$, $j=1,2$
    \FOR{$k=0,1,2,...$}
    \STATE Collect the trajectory $\tau_t = (\textbf{S}_t, a_t, DSR_t, \textbf{S}_{t+1})_{t=0}^{T-1}$ by allocating the policy in Equation \ref{eq: 13}
    \STATE Compute advantages $\hat{A}_t$ by current value $V_{\omega_t}(\textbf{S}_t)$
    \STATE Compute the policy ratio $ \frac{\pi_{\theta_t}(a_j|\textbf{S}_t)}{\pi_{\theta_{t-1}}(a_j|\textbf{S}_t)}$
    \STATE Compute the loss $\mathcal{L}^{OT}$ and $\mathcal{L}^{dis}$ in Equation \ref{eq: 15}
    \STATE Update the policy network by maximizing the clipped objective using $\mathcal{L}^{actor}(\theta)$ in Equation \ref{eq: 15} (both for actor\ 1 and actor\ 2)
    \STATE Update the critic network by minimizing loss $\mathcal{L}^{VF}(\omega)$ in Equation \ref{eq: 9}
    \ENDFOR
\end{algorithmic}
\label{alg:DOT}
\end{algorithm}
\subsection{Optimal Transport Regularization}\label{44}
However, the model lacks a mechanism to ensure the effective allocation of samples to actors. Sometimes, the majority of samples are assigned to one actor.
We incorporate OT techniques to ensure that the Allocation Module assigns more appropriate samples to each actor, thereby capturing diverse patterns more accurately.

We need to consider two main requirements. Firstly, the Allocation Module should allocate the samples to the actor with the smallest decision error. In other words, if $|a_{teacher\ t}^i-a_t^{i1}| > |a_{teacher\ t}^i-a_t^{i2}|$, we tend to assign the sample to actor 2. Secondly, the allocation of samples to the actors should be proportional to their respective patterns. 

Below, we formally define the allocation problem. Assume we utilize $N$ samples in each epoch of PPO's gradient descent process. Based on the error vector, we can construct an error matrix denoted as $L_{err}\in [N\times2]$. Each element $L_{err}^{ij}$ in it represents the decision error of the i-th sample on the j-th actor, given by $L_{err}^{ij} = a_{teacher}^i - a^{ij}$. Corresponding to $L_{err}$ is the allocation matrix $M\in [N\times2]$, where each element $M^{ij}\in \left \{0, 1\right\}$. The value of 1 in the allocation matrix $M$ indicates that Allocation Module assigns the i-th sample to the j-th actor, while the value of 0 indicates no allocation.

The OT method is particularly suitable for solving allocation problem. OT 
involves determining an optimal allocation of resources from one location to another while minimizing overall cost or distance. It is also commonly employed to measure the difference between two probability distributions. Our research aims to find the optimal allocation scheme that minimizes $L_{err}$. The specific formulation of the problem is as follows,
\begin{equation}
\begin{aligned}
    \mathop{min}\limits_{M}& \ (L\cdot M) \\
    s.t.& \frac{\sum_{i=1}^{N} M^{i1}}{N}=w_1,\ \frac{\sum_{i=1}^{N} M^{i2}}{N}=w_2,\ M^{i1}+M^{i2}=1, \forall i= 1,2, ...\ ,N,\\
\end{aligned}
\end{equation}
where $w_1$ and $w_2$ represent the proportions corresponding to different modes (assumed to be $\frac{1}{2}$). We employ the Sinkhorn method to solve the OT problem \cite{cuturi2013sinkhorn}. Figure \ref{fig: OT} provides a visual explanation of the problem we aim to address.

To align the distribution of the output $\textbf{q}^i$ from the allocation module with $M^i$ of the OT problem, we incorporate a cross-entropy loss term. Considering Allocation Module as part of actors, Equation \ref{eq: 11} can be expanded to Equation \ref{eq: 15}, $\lambda_O$ is the hyperparameter.
The third term is $L^{OT}$. 
The pseudocode for the \themodel is shown in Algorithm \ref{alg:DOT}.
 \begin{equation}
\label{eq: 15}
\mathcal{L}^{actor}(\theta)=\mathcal{L}^{CLIP}(\theta)+ \mathcal{L}^{dis}+\lambda_O \sum_{k=1}^{2}M_t^{ik}log(\textbf{q}_t^{ik}).
\end{equation}

\section{Experiments}

\subsection{Dataset}
We utilize the IF stock index futures dataset whose underlying asset is the CSI 300 Index.
The dataset provides minute-level trading data of contracts. Each minute bar includes OHLC, trading volume, etc. The total trading duration in a day is 240 minutes.
We collected it from ricequant.com\footnote{A well-known Chinese quantitative trading platform, https://www.ricequant.com/.}, and divided the data into a training set from 2015-12-31 to 2018-05-08 and a test set from 2018-05-09 to 2019-05-09.


\subsection{Baselines, Evaluation Metrics and Hyperparameters}
\textbf{Baselines}: Long Position Hold (buy futures and hold), Short Position Hold (borrow contracts and hold), Dual Thrust~\cite{kim2007hybrid} (a technical analysis trading strategy commonly used for intraday trading), GRU~\cite{chung2014empirical} (a variant of RNNs\footnote{We chose it as a baseline because we employed the GRU method in the Pretrain Module before imitation learning. The results of GRU demonstrate the performance of the Pretrain Module.}), PPO~\cite{schulman2017proximal} (a RL method that improves stability by preventing large policy changes\footnote{We enhance PPO  using imitation learning mentioned in Methodology Section.}), iRDPG~\cite{liu2020adaptive} (SOTA: an off-policy algorithm that incorporates expert strategy and behavior cloning). 

\noindent\textbf{Evaluation Metrics}: We will measure the model’s performance by Accumulated Rate of Return (ARR, the overall profitability), Volatility (VO, measures by standard deviation of profit $r$), Annualized Sharpe Ratio (ASR, annualized version of $SR$), Maximum Drawdown (MDD, the maximum decline of an asset's value from its peak to the lowest over a period), Calmar Ratio (CR=$\frac{ARR}{MDD}$, risk-adjusted ARR based on MDD) and Sortino Ratio (SoR=$\frac{mean(r)}{std(min(r, 0))}$, excess return per unit of downside risk). 
\noindent\textbf{Hyperparameters}: We set transaction fee rate $\mu =2.3 \times 10^5$ and slippage $\sigma= 0.2$. 
Insufficient account assets may trigger a forced liquidation. We set the margin threshold as $70\%$ and initial capital $C=50000\ CNY$. We repeated 6 experiments for each model.

\begin{table}[tb]
  \caption{Experimental Results ($\uparrow$ indicates the higher the better, $\downarrow$ indicates the opposite)}
  \label{tab:experimental}
  \begin{tabular}{ccccccc}
    \toprule
    \textbf{Methods} & \textbf{ARR ($\uparrow$)} & \textbf{VO ($\downarrow$)} & \textbf{ASR ($\uparrow$)} & \textbf{MDD ($\downarrow$)} & \textbf{CR ($\uparrow$)} & \textbf{SoR ($\uparrow$)}\\
    \midrule
    Long Hold           &$-2.598$    &$0.261$&$-0.638$         &$113.121$   &$-0.001          $&$-0.080$\\
    Short Hold          &$3.163$     &$0.259$         &$0.782$          &$0.894$          &$0.041           $&$0.093 $\\
    Dual Thrust         &$10.130$      &$0.253$        &$2.628$          &$0.033$          &$3.962           $&$0.365 $\\
    GRU                 &$11.342(1.12)$&$0.242(0.00)$&$3.004(0.31)$&$0.016(0.02)$&$4.280(0.23)$&$0.399(0.05)$\\
    iRDPG               &$14.453(0.98)$&$0.254(0.01)$&$3.955(0.18)$&$0.023(0.03)$&$5.881(3.21)$&$0.537(0.03)$\\
    PPO                 &$12.245(0.23)$&$0.243(0.00)$&$3.223(0.05)$&$0.022(0.02)$&$4.281(0.23)$&$0.436(0.01)$\\
    \themodel-ND        &$15.322(1.25)$&$0.246(0.01)$ &$4.252(0.24)$	&$\textbf{0.005}(\textbf{0.01})$	&$\textbf{7.277}(\textbf{3.51})$	&$0.587(0.07)$ \\
    \themodel-NO        &$17.236(1.05)$&$0.248(0.01)$	&$4.447(0.18)$	&$0.026(0.01)$	&$5.558(0.75)$	&$0.529(0.08)$\\
    \textbf{\themodel}  &$\textbf{20.379} (\textbf{0.85})$&$\textbf{0.228} (\textbf{0.00})$&$\textbf{ 5.395}  (\textbf{0.26})$&$ 0.011(0.02)$&$6.582(0.66)$&$\textbf{0.605}  (\textbf{0.05})$\\
    \bottomrule
  \end{tabular}
\end{table}
\begin{figure}[tb]
  \includegraphics[width=\textwidth]{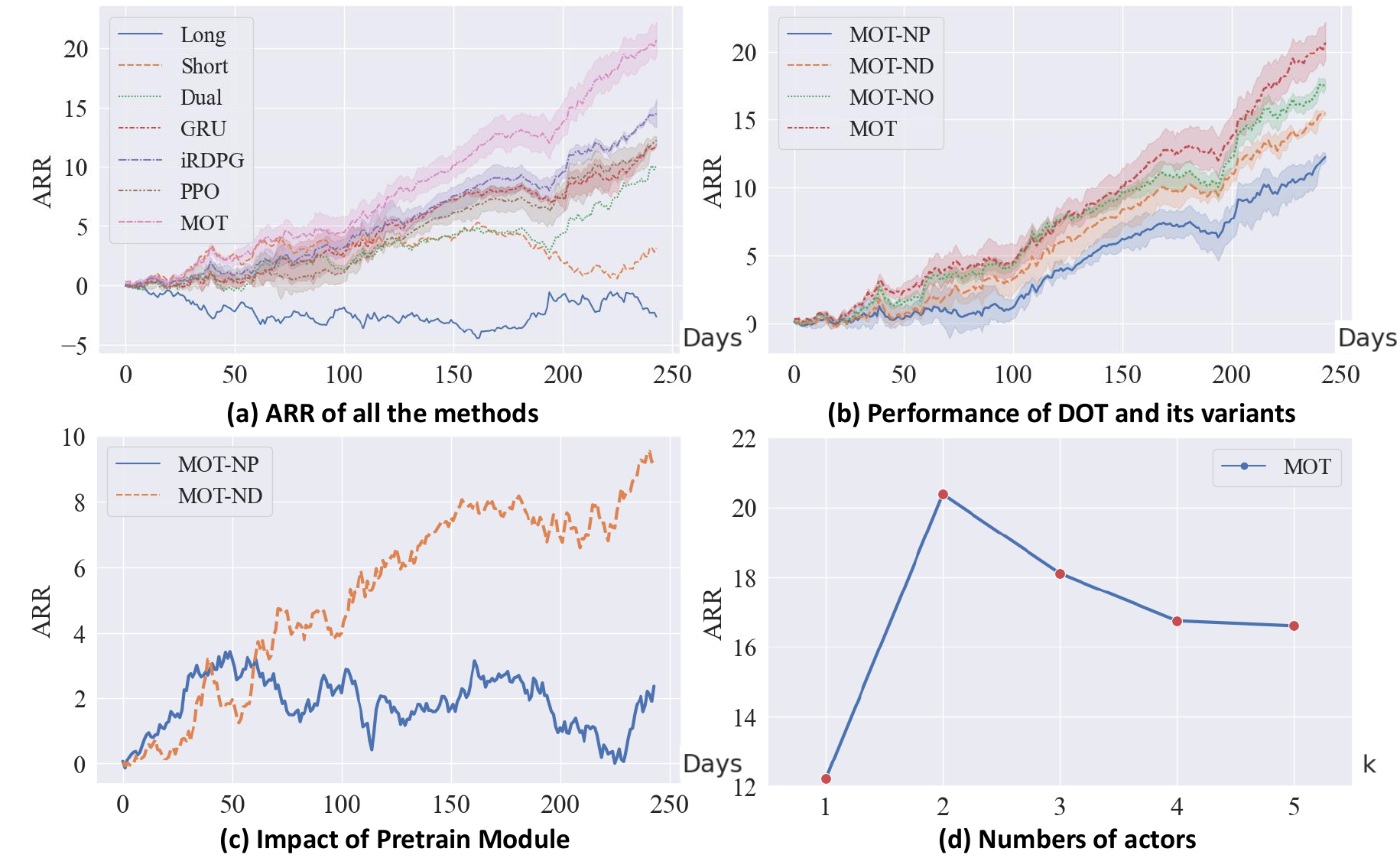}
  \caption{Performance of different models in terms of ARR}
  \label{fig: ARR}
\end{figure}
\subsection{Experimental Results}
Table \ref{tab:experimental} provides a summary of the results. Figure \ref{fig: ARR} (a) depicts ARR of all the methods. From Table 2, \themodel outperforms other methods in terms of profit and risk-reward balance. ARR is the most crucial indicator, and our model achieves the highest ARR. The ARR of PPO is about 1.0 higher than that of GRU, indicating that PPO exhibits greater robustness. 
The ASR, CR, and SoR are composite metrics that consider both risk and return. Deep learning methods (last 6 rows in Table \ref{tab:experimental}) outperform the technical indicator models (first 3 rows in Table \ref{tab:experimental}) in these three metrics, which suggests the former better represents complex states under high-noise conditions.
\themodel performs second in terms of MDD, indicating that \themodel only requires a short time period to recover from losses.
RL models outperform time-series models, as the latter primarily focuses on predicting price trends without considering the high costs caused by incorrect predictions. 
Since greater risk leads to greater returns, profits are higher when there are significant price fluctuations. So the correlation among most methods is very high.

\subsection{Ablation Study}
We conducted ablation experiments to show the effectiveness of its three components. The experimental results and the trend of ARR are depicted in Table \ref{tab:experimental} and Figure \ref{fig: ARR} (b).

\textbf{Overall performance. }\themodel-NP applies imitation learning based on PPO without Pretrain Module. \themodel-ND is obtained by removing multiple actors from the final model, while \themodel-NO eliminates the process of OT. From Figure \ref{fig: ARR} (b), we observe that ARR curve of \themodel remains higher than other variants in most periods. Table \ref{tab:experimental} shows that \themodel performs best in terms of ARR, VO, ASR, and SOR. Among the three modules, OT method contributes the most to the improvement of model performance, followed by the Pretrain Module. \themodel-ND excels in MDD metric, indicating that the model without multiple actors' design tends to generate more conservative strategies. While a conservative trading strategy often misses the optimal investment opportunities. 
Since the calculation of CR relies on MDD, \themodel-ND also exhibits higher CR. 

\textbf{Effectiveness of Pretrain Module. }The influence of the expert strategy in DB diminishes over time and the benefit of imitation learning is mainly observed in the early stages. For the ablation experiment, we selected the agent trained for 100 epochs after imitation learning. Figure \ref{fig: ARR} (c) illustrates the impact of Pretrain Module on imitation learning and the yellow curve is the model with Pretrain Module. It can be observed that \themodel-ND demonstrates a steady increase accompanied by minor fluctuations in profit. In contrast, \themodel-NP experiences some declines and doesn't learn well. This indicates that Pretrain Module contributes to the improvement of imitation learning.

\begin{figure}[t]

  \centering
  \includegraphics[width=\linewidth]{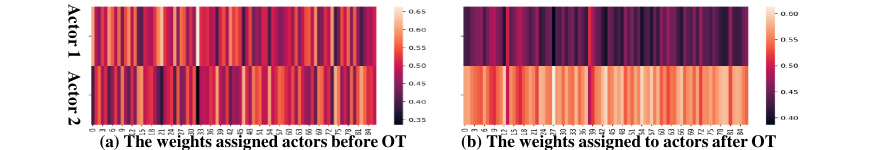}
  \caption{Effectiveness of OT modeling}
  \label{fig: heat}
\end{figure}

\textbf{Effectiveness of multiple actors and OT modeling. }Figure \ref{fig: heat} demonstrates the variation in weights assigned to two actors before and after 
OT modeling. In a relatively volatile period, the model assigns weights more randomly without OT
while assigns higher weights to actor 2 with OT. Notably, the introduction of OT leads to higher returns and enhances the ability to capture complex patterns. 
Figure \ref{fig: ARR} (d) illustrates the impact of actors' number to \themodel. \themodel achieves the best profitability when $k=2$ while achieves the worst when $k=1$. This indicates that only one actor is insufficient to capture all patterns, while an excessive number of actors may lead to redundancy. In our model, the optimal number of actors is 2.
\section{Related Work}

\textbf{Investment strategies based on expert knowledge.} 
The early method used expert knowledge to construct heuristic rules
\cite{jegadeesh1993returns,poterba1988mean}, which can be divided into two categories: fundamental analysis and technical analysis. Fundamental analysis captures diverse factors such as industry trends, company financial statements, and public opinion. This method is more commonly used by long-term investors to find undervalued assets. 
Popular technical indicators include Relative Strength Index \cite{wilder1978new}, Average Direction Index \cite{gurrib2018performance}, On-Balance Volume \cite{tsang2009profitability}
, etc. Commonly used investment strategies include momentum trading \cite{hong1999unified} and mean reversion strategy \cite{poterba1988mean}. However, interrelated technical indicators are correlated with each other, and building them directly from the market introduces too much market noise.
Typically, rules constructed based on expert knowledge
can only capture trading opportunities under specific market conditions \cite{deng2016deep}.

\noindent\textbf{Investment strategies based on RL.} 
In contrast to supervised learning, which still requires expert knowledge to construct strategies, RL can optimize strategies
in an end-to-end form.
Moody et al.~\cite{moody1997optimization} made the first attempt to apply recurrent RL (RRL) algorithm to algorithmic trading.
However, traditional RL methods are not well-suited for environments with large state spaces, making it challenging to select market features. Deep RL methods have partially addressed this problem.
Si et al.~\cite{si2017multi} argue that strategies need to consider multiple factors
and combine multi-objective optimization with deep RL to address this issue. Oliveira et al.~\cite{de2020tabular} adopts SARSA, which maps states and actions to specific cells in a table to learn the value function. Since insufficient financial data causes overfitting, Jeong et al.~\cite{jeong2019improving} divided stocks into groups based on their correlations and introduced transfer learning into the Deep Q-Network (DQN). 
To shorten the inefficient random exploration phase, iRDPG \cite{liu2020adaptive} incorporates technical analysis through imitation learning.
Yuan et al.~\cite{yuan2020using} argue that daily frequency data cannot meet the high demands of RL and instead use minute frequency data. 
And PPO algorithm achieves more stable returns compared to DQN and SAC algorithms.


\section{Conclusion}
In this paper, we propose \themodel, an RL-based model for algorithmic trading problems. Specifically, we model the algorithmic trading problem as MDP and leverage imitation learning to enable the agent to learn from expert knowledge. To better initialize \themodel, we introduce the Pretrain Module prior to the imitation learning phase. Considering that futures prices result from different patterns, we employ multiple actors with disentangled representation learning to model the patterns. We design the Allocation Module to integrate the outputs of multiple actors and incorporate OT techniques to guide the learning of the Allocation Module. Experimental results demonstrate that our model achieves superior profitability while controlling the risk, showcasing its robustness in financial markets with complex data patterns. Further ablation studies confirm the effectiveness of the three components of \themodel.

\subsubsection{Acknowledgements} This work was supported by the National Natural Science Foundation of China (No. 72374201).
\bibliographystyle{splncs04}
\bibliography{chengxi}
\end{document}